\documentclass{svjour3}                     % onecolumn (standard format)
\smartqed  % flush right qed marks, e.g. at end of proof
\usepackage{graphicx}
\usepackage{braket}

\usepackage{amssymb}
\usepackage{tensor}

\begin{document}

\title{Power spectrum for perturbations in an inflationary model for a closed universe}

\titlerunning{}        

\author{Claus Kiefer  \and Tatevik Vardanyan 
}

\institute{C. Kiefer \at
              Institute for Theoretical Physics,
              University of Cologne\\
              Z\"ulpicher Stra\ss e 77,
              50937 K\"oln, Germany\\
              \email{kiefer@thp.uni-koeln.de}  
           \and
T. Vardanyan \at
Institute for Theoretical Physics,
              University of Cologne\\
              Z\"ulpicher Stra\ss e 77,
              50937 K\"oln, Germany
 \\ \email{tatevik@thp.uni-koeln.de} \\ 
}

\date{Received: date / Accepted: date}

\maketitle

\begin{abstract}
We derive the power spectrum of primordial quantum fluctuations in an inflationary universe for curvature parameter ${\mathcal K}=1$. This is achieved through a Born--Oppenheimer type of approximation scheme from the Wheeler--DeWitt equation of canonical quantum gravity using gauge-invariant variables. 
Compared to the flat model, the closed model exhibits a deficit of power at large scales.
\end{abstract}

\vskip 3mm

\noindent {\em Invited contribution to a Topical Collection for the GRG journal in memory of Prof. T. Padmanabhan}

\section{\label{sec:1}Introduction}

Observations seem to indicate that our Universe is spatially flat \cite{paper:Planck2018}. For this reason, one mostly uses the flat case ${\mathcal K}=0$ in Friedmann models for the present and the early Universe. This does, of course, not imply that our Universe is spatially {\em exactly} flat; it only implies that potential curvature terms seem irrelevant in the available class of observations. Still, there is currently a discussion about the correct interpretation of the data, and indications were found that speak in favour of a spatially closed Universe; see, for example, \cite{paper:Valentino}, \cite{paper:Handley,thavanesan}. This is one of the reasons why we undertake here the task of calculating the power spectrum for a spherical (${\mathcal K}=1$)-model. A closed model might explain the observed low amplitude of quadrupole and octopole modes \cite{paper:Planck2018}, which remains unexplained in the standard $\Lambda{\rm CDM}$ model of cosmology. 
In addition, an inflationary phase has been invoked to solve the flatness problem; that is, even for general curvature parameter ${\mathcal K}$ the dynamics of the inflation drives the universe toward flatness, by which we mean a negligible term $\Omega_{\mathcal K}=-{\mathcal K}/a^2H^2$. From this viewpoint, therefore, considering a universe with  curvature (in our case ${\mathcal K}=1$) is in the context of inflation more natural than just assuming exact flatness (${\mathcal K}=0$).

Another reason for our investigation is to have the formalism ready for future applications in quantum gravity. The closed case was, of course, addressed before,\footnote{See, for example, \cite{papers:closed,closed2,closed3,closed4}.} but it is, to our knowledge, the first time that a full analysis from a fundamental (Schr\"odinger) picture point of view is performed using gauge-invariant variables. 

Studying a closed universe is not restricted to the spherical case. One could also envisage a three-dimensional torus universe or a topologically more involved model. To be definite, we shall discuss here only the case of a three-dimensional sphere. 

In our paper, we employ cosmic perturbation theory for an inflationary universe; see, for example, \cite{paper:Mukhanov}, \cite{book:Paddy1993}, and \cite{book:Peter}. Since in such models the perturbations have their origin in primordial {\em quantum} fluctuations, we find it natural to employ the Schr\"odinger picture for their formulation. For the flat case, such a description was used, for example, in \cite{book:Paddy1993}, \cite{paper:Martin}, and \cite{paper:David,paper:Manuel}. As for the initial state of these quantum fluctuations, we make the standard assumption that they be in their adiabatic vacuum state, an assumption that may eventually be justified from a fundamental principle such as the (quantum) Weyl curvature hypothesis \cite{paper:Kiefer2021}. The extension to excited states is straightforward \cite{paper:excited}. 

Our paper is organized as follows. In section~2, we review the cosmic perturbation theory for a spherical universe. We use gauge-invariant variables, in particular a generalization of the Mukhanov--Sasaki variable known from the flat case. Curvature perturbations are introduced in section~3. In section~4, we discuss the quantization of the modes in the slow-roll approximation using a Born--Oppenheimer type of scheme for the Wheeler--DeWitt equation of canonical quantum gravity. As initial condition we choose the modes to be in their adiabatic ground state. Section~5 is then devoted to the calculation of the power spectrum. We compare the result with the power spectrum for the flat case and show that there is less power for large modes. We end with a brief conclusion and add some appendices displaying details of the perturbation theory for the closed case. 

\section{\label{sec:2}Hamiltonian formulation for an inflationary Friedmann universe}
The construction of a Hamiltonian formulation for general relativity  starts with the foliation of spacetime into three-dimensional spacelike Cauchy hypersurfaces parameterized by a time function $t$ (see e.g. \cite{book:Kiefer}).
The line element takes the form
\begin{equation}
    d s^2=-(N^2-N_i N^i)d t^2+2N_i d x^i d t +h_{ij} d x^i d x^j,
    \label{ADM}
\end{equation}
where $N$ is the lapse function, $N_i$ is the shift vector, and $h_{ij}$ is the three-metric.

The total action, which is a sum of the Einstein--Hilbert action and the action of a scalar field, will be rewritten using (\ref{ADM}). Afterwards, taking variations of the action with respect to the lapse function and shift vector leads to the Hamiltonian and momentum constraints. For the 
Friedmann--Lema\^{\i}tre--Robertson--Walker (FLRW) metric, see (\ref{FLRW}) below,
the momentum constraints are trivially satisfied.

\subsection{\label{subsec:21}Background Hamiltonian}
To describe an inflationary universe, we consider the FLRW metric coupled to a scalar field $\phi$ with a quadratic potential $\mathcal{V}(\phi)=\frac{1}{2}m^2\phi^2$.
For a closed spherical universe, the line element reads
\begin{equation}
    ds^2=-N^2(t)dt^2+a^2(t)d\Omega_3^2,
    \label{FLRW}
\end{equation}
where $d\Omega_3^2$ denotes the metric of the unit three-sphere. We note that the scale factor $a$, the Friedmann time $t$ (and thus $s$) have dimensions of a length, while the remaining variables are dimensionless.

Redefining the scalar field as $\phi \rightarrow \phi/\sqrt{2} \pi$, the action for the inflationary universe takes the form
\begin{equation}
        S=\frac12\int N\ d t\Bigg[ -\frac{3\pi}{2G}\frac{a}{N^2}{\dot{a}}^2+\frac{a^3}{N^2}{\dot{\phi}}^2+\frac{3{\pi}a\mathcal{K}}{2 G}-2a^3\mathcal{V}(\phi)\Bigg],
        \label{EHa}
\end{equation}
where the spatial integration was performed over the three-sphere.

Let us introduce the variable $m_{\rm P}:=\sqrt{\frac{3\pi}{2 G}}$, which in the quantum theory will become the (redefined) Planck mass after setting $\hbar=1$.
The background Hamiltonian can then be written in the form
\begin{equation}
 H_{|0}=N\frac12 e^{-3\alpha}\Big[-\frac{1}{m_{\rm P}^2 }{p_{\alpha}}^2+{p_{\phi}}^2+2e^{6\alpha}\mathcal{V}(\phi)-m_{\rm P}^2\mathcal{K}e^{4\alpha }\Big],
 \label{H0}
\end{equation}
where
\begin{equation}
p_{\alpha}=-m_{\rm P}^2e^{3\alpha}\frac{\dot{\alpha}}{N}\ \ \ \textnormal{and} \ \ \ p_{\phi}=e^{3\alpha}\frac{\dot{\phi}}{N}.
\label{P}
  \end{equation}
We have introduced here the dimensionless quantity $\alpha$ by $\alpha=\ln{a/a_0}$, where $a_0$ is a reference scale that we shall not write out explicitly (one should keep in mind that it occurs together with powers of $e^{\alpha}$).
The field equations are
\begin{eqnarray}
 H^2&=&\frac{1}{m_{\rm P}^2}\Big(2\mathcal{V}(\phi)+{\dot{\phi}}^2\Big)- \frac{\mathcal{K}}{a^2},\label{F1}\\
 \frac{{\ddot{a}}}{a}&=&\frac{1}{m_{\rm P}^2}\Big(2\mathcal{V}(\phi)-2{\dot{\phi}}^2\Big)\label{F2},\\
 {\ddot{\phi}}&+&3H{\dot{\phi}}+\mathcal{V}_{,\phi}=0.
 \label{KG}
\end{eqnarray}
Note that in order to write the above expressions in terms of conformal time $\eta$, which is defined by $d\eta/d t=a^{-1}$, one simply has to set $N=a(t)/a_0$. Also, the derivative with respect to $\eta$ will be denoted by a prime. Conformal time will be used below.

Solutions of the classical background equations for closed models and quadratic scalar potential were discussed, for example, in \cite{BGKZ85} and \cite{HW85}. In the inflationary regime, there are no significant differences to the flat case. The main difference lies in the presence of a recollapse in the closed case. This will be different for the perturbations, which are the main subject of our paper, and to which we now turn. 

\subsection{\label{subsec:22}Perturbation Hamiltonian}
Including perturbations to the metric as well as to the scalar field (see details in Appendix~\ref{app:sec2}), the total Hamiltonian can be given as a sum of the background and perturbation parts \cite{paper:Hawking},
\begin{equation}
  H=N\Big[H_{|0}+\sum_{n}{H_{|2}^n} \Big].
  \label{H}
\end{equation}
The perturbation part ${H_{|2}^n}$ is composed of scalar, vector, and tensor parts,
\begin{equation}
{H_{|2}^n}=\sum_n\Big[ {^S{H_{|2}^n}}+ {^V{H_{|2}^n}}+ {^T{H_{|2}^n}}\Big],\\
\label{Hn}
\end{equation}
where only the scalar part ${^S{H_{|2}^n}}$ is of interest here; therefore, the vector and tensor parts will be ignored.
The first-order Hamiltonians
\begin{equation}
    H^n_{\ |1}=0, \ \ \ {^S{H^n_{\ \_1}}}=0,\ \ \ {^V{H^n_{\,\_1}}}=0
    \label{constraints}
\end{equation}
are the linearized Hamiltonian constraint and the scalar and vector parts of the linearized momentum constraints, respectively. Explicit expressions can be found in \cite{paper:Hawking}; the structure of the perturbation scheme is summarized in Sec.~II of \cite{paper:David}. 

\subsubsection{\label{subsubsec:221}Construction of gauge-invariant variables}

The Hamiltonian (\ref{H}) obtained by Halliwell and Hawking in \cite{paper:Hawking} is not gauge-invariant. A gauge-invariant form  was obtained by Langlois in \cite{paper:Langlois}. Applying a method developed for isolating physical degrees of freedom for Hamiltonian systems with a set of first-class constraints \cite{paper:Goldberg}, he obtained gauge-invariant variables and presented the scalar part ${^s{H_{|2}^n}}$ of the perturbation Hamiltonian in gauge-invariant form. 
This gauge-invariant variable reads
\begin{equation}
    Q_n:=-\frac{\phi^\prime}{\mathcal{H}}(a_n+b_n)+f_n,
    \label{Qn}
\end{equation}
where $a_n$, $b_n$, and $f_n$ are the expansion coefficients defined in Appendix~B, following \cite{paper:Hawking}. 
The gauge-invariant perturbation Hamiltonian assumes the form
\begin{equation}
    {^s{H_{|2}^n}}=\frac12 N \Big\{e^{-3\alpha}U_n(\alpha,\phi)P_n^2+W_n(\alpha,\phi)Q_n^2\Big\},
    \label{HnQn}
\end{equation}
where
\begin{equation}
    U_n(\alpha,\phi)=1+\frac{3m_{\rm P}^2\mathcal{K}}{n^2-4\mathcal{K}}\Bigg(\frac{p_{\phi}}{p_{\alpha}}\Bigg)^2,
    \label{Un}
\end{equation}
and
\begin{eqnarray}
 W_n(\alpha,\phi)=&e^{\alpha}&(n^2-\mathcal{K})+e^{3\alpha}m^2 +\frac{1}{n^2-4\mathcal{K}}\frac{1}{U_n^2}\Bigg[\frac{3}{m_{\rm P}^2}(n^2-\mathcal{K})\Bigg(1-\frac{n^2-7\mathcal{K}}{n^2-4\mathcal{K}}\Bigg)e^{-3\alpha}{p_{\phi}}^2  \nonumber
 \\&+&2m_{\rm P}^2e^{9\alpha}\frac{m^4\phi^2}{{p_{\alpha}}^2}-9m_{\rm P}^2\frac{n^2-\mathcal{K}}{n^2-4\mathcal{K}}e^{-3\alpha}\frac{{p_{\phi}}^6}{{p_{\alpha}}^4} -2(2n^2-5\mathcal{K})e^{3\alpha}m^2\phi\frac{p_{\phi}}{p_{\alpha}}\nonumber
 \\&-&6m_{\rm P}^2e^{3\alpha}m^2\phi{\Bigg(\frac{p_{\phi}}{p_{\alpha}}\Bigg)}^3-m_{\rm P}^2e^{3\alpha}m^2{\Bigg(\frac{p_{\phi}}{p_{\alpha}}\Bigg)}^2-6m_{\rm P}^2e^{9\alpha}m^4\phi^3\frac{p_{\phi}}{p_{\alpha}^3} \nonumber
 \\&+&4m_{\rm P}^4\mathcal{K}e^{7\alpha}m^2\phi\frac{p_{\phi}}{p_{\alpha}^3}-\frac{3m_{\rm P}^4 \mathcal{K}}{n^2-4\mathcal{K}}e^{3\alpha}m^2{\Bigg(\frac{p_{\phi}}{p_{\alpha}}\Bigg)}^4+3(n^2-\mathcal{K})e^{3\alpha}m^2\phi^2{\Bigg(\frac{p_{\phi}}{p_{\alpha}}\Bigg)}^2   \nonumber
 \\&-&2m_{\rm P}^2\mathcal{K}e^{\alpha}(n^2-\mathcal{K}){\Bigg(\frac{p_{\phi}}{p_{\alpha}}\Bigg)}^2-9m_{\rm P}^2\frac{n^2-\mathcal{K}}{n^2-4\mathcal{K}}e^{3\alpha}m^2\phi^2{\Bigg(\frac{p_{\phi}}{p_{\alpha}}\Bigg)}^4  \nonumber
 \\&+&6m_{\rm P}^4\mathcal{K}e^{\alpha}\frac{n^2-\mathcal{K}}{n^2-4\mathcal{K}}{\Bigg(\frac{p_{\phi}}{p_{\alpha}}\Bigg)}^4\Bigg].
  \label{Wn}
\end{eqnarray}
For the large-wavelength limit, i.e. $n\rightarrow \infty$, the perturbation Hamiltonian (\ref{HnQn}) reduces to
\begin{eqnarray}
      {^s{H_{|2}^{n}}}=\frac N2\Bigg\{e^{-3\alpha}P_n^2+\Bigg[&e^{\alpha}n^2&+e^{3\alpha}m^2+\frac{18}{m_{\rm P}^2}e^{-3\alpha}{p_{\phi}}^2\nonumber\\&-&18e^{-3\alpha}\frac{{p_{\phi}}^4}{{p_{\alpha}}^2}-12e^{3\alpha}m^2\phi\frac{p_{\phi}}{p_{\alpha}}\Bigg]Q_n^2\Bigg\},
      \label{Hnflat}
\end{eqnarray}
which corresponds to the result for the flat case \cite{paper:Langlois}. We note the occurrence of momenta $p_{\alpha}$ in the denominator, which can present problems for the general quantization, but remains unproblematic for the semiclassical approximaton scheme employed below. 

Using the Hamilton equations
\begin{eqnarray}
    \dot{Q}_n&=&\frac{\partial ({^s{H_{|2}^{n}}} )}{\partial P_n}=N e^{-3\alpha}U_n(\alpha,\phi)P_n,    \label{Hameq1}\\
    \dot{P}_n&=&-\frac{\partial ({^s{H_{|2}^{n}}})}{\partial Q_n}=-N e^{-3\alpha}W_n(\alpha,\phi)Q_n,
    \label{Hameq2}
\end{eqnarray}
we obtain the equations of motion for the gauge-invariant variable $Q_n$,
\begin{equation}
{\ddot{Q}}_n+b(n,\alpha,\phi){\dot{Q}}_n+c(n,\alpha,\phi)Q_n=0,
\label{eomQn}
\end{equation}
where
\begin{equation}
b(n,\alpha,\phi)=1-\frac{\dot{N}}{N}-\frac{\dot{U}_n(\alpha,\phi)}{U_n(\alpha,\phi)}\quad \textnormal{and} \quad
c(n,\alpha,\phi)=N^2e^{-3\alpha}U_n(\alpha,\phi)W_n(\alpha,\phi).
\end{equation}

\subsubsection{\label{subsubsec:222}Canonical transformation}
We now perform a canonical transformation $(Q_n,P_n)\rightarrow(\tilde{v}_n,\tilde{p}_n)$ to get a more compact form for the perturbation Hamiltonian (\ref{HnQn}). Moreover, we will use the conformal time $\eta$ in the following calculations.  
We use the generating function $F_3(P_n,\tilde{v}_n,\eta)$, which is a function of the old momenta and the new coordinates.

Using the relations
\begin{eqnarray}
    Q_n&=-\frac{\partial F_3}{\partial P_n}, \label{F31}\\
    \tilde{p}_n&=-\frac{\partial F_3}{\partial \tilde{v}_n},\label{F32}
\end{eqnarray}
the new Hamiltonian in terms of $(\tilde{v}_n,\tilde{p}_n)$ is obtained via
\begin{equation}
    H_{\rm new}=H_{\rm old}+\frac{\partial F_3}{\partial \eta}.
    \label{Hnew}
\end{equation}
We now apply this transformation scheme to our particular case. We relate the new variables $(\tilde{v}_n,\tilde{p}_n)$ and the old variables $(Q_n,P_n)$ by
\begin{eqnarray}
 {\tilde{v}_n}&=& \frac{a }{\sqrt{U_n}}Q_n,\label{tildev}\\
 \tilde{p}_n&=& \frac{\sqrt{U_n}}{a} P_n +\left(\mathcal{H}-\frac12\frac{U_n^{\prime}}{U_n}\right)\tilde{v}_n,\label{tildep}
\end{eqnarray}
where a dimensionless Hubble parameter $\mathcal{H}$ is defined by $\mathcal{H}=a^\prime/a$. Let us call the new variable $\tilde{v}_n$ a generalized Mukhanov--Sasaki variable.
From this, we obtain the generating function
\begin{equation}
    F_3=-g^{-1}{\tilde{v}_n}p_n-\frac{g^{\prime}}{g}\frac{{\tilde{v}_n}^2}{2}. 
    \label{F3}
\end{equation}
Substituting the time derivative of the generating function into (\ref{Hnew}), we find the Hamiltonian in terms of the generalized Mukhanov--Sasaki variable $\tilde{v}_n$ and its conjugate momentum $\tilde{p}_n$,
\begin{equation}
{^s{H_{|2}^n}}=\frac12\Big[{\tilde{p}_n}^2+\omega_n^2(\eta){\tilde{v}_n}^2  \Big],
\label{Hnnew}
\end{equation}
where 
\begin{equation}
\omega_n^2(\eta)=\frac{1}{a}W_n \,U_n-{{\mathcal{H}}^2} -{\mathcal{H}^{\prime}}+\frac{\mathcal{H}U_n^{\prime}}{U_n}-\frac34\frac{{U_n^{\prime}}^2}{{U_n}^2}+\frac12\frac{U_n^{\prime \prime}}{U_n}.
\label{omega}
\end{equation}
In the large-wavelength limit we have $\tilde{v}_n=a Q_n$, and the perturbation Hamiltonian (\ref{Hnflat}) assumes the form (\ref{Hnnew}) with
\begin{equation}
   \omega_n^2=n^2-\frac{z^{\prime \prime}}{z}, \quad \textnormal{and} \quad z=\frac{a\phi^\prime}{\mathcal{H}}.
   \label{omegaflat}
\end{equation}
This corresponds to the well-known result for the flat case \cite{paper:Martin}.

\section{\label{sec:3}Curvature perturbations}
Let us introduce now a gauge-invariant variable $\zeta_{\rm BST}$\footnote{Named after Bardeen, Steinhardt, and Turner \cite{paper:BST}.} which can be written in terms of a gauge-invariant metric potential $\Phi$ defined below in (\ref{Psi}), as follows:\footnote{See, for example, \cite{book:Peter}, Eq.~(5.154), or \cite{paper:Schwarz}, Eq.~(4.12).}
\begin{equation}
  \zeta_{\rm BST}=-\frac23\frac{{\mathcal{H}}^2}{(1+w)({\mathcal{H}}^2+\mathcal{K})}\Bigg\{{\mathcal{H}}^{-1}\Phi^{\prime}+\Bigg[1-\frac{\mathcal{K}}{{\mathcal{H}}^2}+\frac13\Bigg(\frac{k}{\mathcal{H}}\Bigg)^2\Bigg]\Phi\Bigg\}-\Phi,
  \label{zetaBST}
\end{equation}
where $w$ is the barotropic index in the equation of state $p=w\rho$. We emphasize that $\zeta_{\rm BST}$ is a conserved quantity during the evolution for super-Hubble modes ($k\gg\mathcal{H}$). Due to this property, it plays an important role in relating the power spectrum of scalar perturbations at the end of inflation to the temperature anisotropies of the CMB. 
 
In the literature, another parameter $\zeta$ is usually taken as a conserved quantity; here, however, it does not serve this purpose because it is conserved only for the spatially flat case.
The two parameters are related as\footnote{See, for example, \cite{book:Peter}, Eq.~(5.156).}
\begin{equation}
    \zeta=\zeta_{\rm BST}-\frac13\frac{\Delta \Phi}{\mathcal{H}^{\prime}-\mathcal{H}^2},
    \label{zeta}
\end{equation}
and for the super-Hubble modes we have $\zeta \simeq \zeta_{\rm BST}$.

Let us express now $\zeta_{\rm BST}$ in terms of $\tilde{v}_n$. This will allow us later to obtain the power spectrum of $\zeta_{\rm BST}$ from the spectrum of $\tilde{v}_n$.
First, we expand $\zeta_{\rm BST}$ and $\Phi$ in scalar harmonics as $X=\sum_{n,l,m}6^{-1/2}X_n Q^n_{lm}$.\footnote{The purpose of having $6^{-1/2}$ in the expansion is to cancel out this extra multiplier from the formulas later on which arises due to the expansion (\ref{epsilon}).} Next, using relations (\ref{Qn}), (\ref{Un}) and (\ref{P}), we write the explicit form of $\tilde{v}_n$ defined by (\ref{tildev}),
\begin{equation}
    \tilde{v}_n=\frac{a}{\sqrt{U_n}}Q_n=a \left[1+\frac{3\mathcal{K}}{n^2-4\mathcal{K}}\bigg(\frac{{\phi}^{\prime}}{\mathcal{H}}\bigg)^2\right]^{-1}\left(-\frac{\phi^{\prime}}{\mathcal{H}}(a_n+b_n)+f_n\right).
    \label{tildevn}
\end{equation}
Since $\zeta_{\rm BST}$ is expressed in terms of the potential $\Phi$, we have to relate functions $a_n$, $b_n$ and $f_n$ to it as well. 
For that purpose, let us recall the line element for a FLRW metric plus scalar perturbations as described by functions $A$, $B$, $C$, $E$ (see e.g. \cite{paper:Mukhanov}),
\begin{equation}
    d s^2=a^2(\eta)\big\{-(1+2A)d\eta^2+2B_{|i} d x^i d\eta+\big[(1+2 C) \gamma_{ij}+2 E_{|ij}\big]d x^i d x^j\big\}.
    \label{pertmet}
\end{equation}
where $\gamma_{ij}$ is the metric on the unit three-sphere.

Combining the above-mentioned functions, one can construct the following two gauge-independent quantities,
\begin{eqnarray}
    \Psi &\equiv& -C-\mathcal{H}(B-E^{\prime}),\label{Psi}\\
    \Phi &\equiv& A+\mathcal{H}(B-E^{\prime})+(B-E^{\prime})^{\prime},
    \label{Phi}
\end{eqnarray}
which are known as Bardeen potentials.
Using the relations given in Appendix~\ref{app:sec2}, where the perturbations of the metric are expanded in scalar harmonics, the line element of the metric plus scalar perturbations can be written as
\begin{eqnarray}
    d s^2=a^2(\eta)\Bigg\{&-&\Big(1+2\cdot 6^{-1/2}\sum_{n,l,m}g_{nlm}Q^n_{lm}\Big)d \eta^2\nonumber\\&+&2\sum_{n,l,m}6^{-1/2}k_{nlm}\frac{1}{(n^2-1)}(Q_{|i})^n_{lm} d x^i d \eta\nonumber \\&+& \bigg[\Big(1+\sum_{n,l,m}6^{1/2}(a_{nlm}+b_{nlm})\frac13Q_{lm}^n\Big)\gamma_{ij}\nonumber\\&+&\sum_{n,l,m}6^{1/2}b_{nlm}\frac{1}{(n^2-1)}(Q_{|ij})^n_{lm}\bigg] d x^i d x^j\Bigg\}.
    \label{pertmetsh}
\end{eqnarray}
%where we have absorbed $6^{-1/2}$ into scalar harmonics $Q^n_{lm}$.
Directly comparing (\ref{pertmet}) and (\ref{pertmetsh}) we can relate the functions $A$, $B$, $C$, $E$ with the functions $a_n$, $b_n$, $g_n$, $k_n$.

For convenience, we will proceed in longitudinal gauge ($B=E=0$), for which the line element reads
\begin{equation}
    d s^2=a^2(\eta)\big\{-(1+2\Phi)d\eta^2+ (1-2\Phi)\gamma_{ij}d x^i d x^j\big\}.
    \label{longmet}
\end{equation}
Given the fact that the two Bardeen potentials (\ref{Psi}) and (\ref{Phi}) are equal \cite{paper:Mukhanov}, \cite{paper:Schwarz}, the following relations are obtained,
\begin{equation}
\Phi_n=g_n=-a_n   
\label{r1}
\end{equation}
and
\begin{equation}
  b_n=k_n=0.  
  \label{r2}
\end{equation}
We can then express the function $f_n$ in terms of $\Phi_n$ by (\ref{fn}); see Appendix~\ref{app:sec3} for details. 

Finally, substituting (\ref{QnPhi}) into (\ref{tildevn}), we arrive at 
\begin{equation}
    \tilde{v}_n=\frac{a\mathcal{H}}{3\phi^{\prime}}\Bigg\{\mathcal{H}^{-1}\Phi_n^{\prime}+\Bigg[1+\frac{3{\phi^{\prime}}^2}{\mathcal{H}^2}\Bigg]\Phi_n\Bigg\}\Bigg[{1+\frac{\mathcal{K}}{n^2-4\mathcal{K}}\frac{3{{\phi}^{\prime}}^2}{\mathcal{H}^2}}\Bigg]^{-1/2}.
    \label{tildevnphi}
\end{equation}
Thus, $ \zeta_{{\rm BST},n}$ and $\tilde{v}_n$ are related by
\begin{equation}
 \zeta_{{\rm BST},n}=-\frac{\mathcal{H}}{a{\phi^{\prime}}}\Bigg[{1+\frac{\mathcal{K}}{n^2-4\mathcal{K}}\frac{3{{\phi}^{\prime}}^2}{\mathcal{H}^2}}\Bigg]^{1/2}\tilde{v}_n-\frac{\mathcal{H}^2}{3{\phi^{\prime}}^2}\frac{(n^2-3\mathcal{K})}{\mathcal{H}^2}\Phi_n,
 \label{zetaBSTphi}
\end{equation}
for which the following relation has been used,
\begin{equation}
 -\frac23\frac{{\mathcal{H}}^2}{(1+w)({\mathcal{H}}^2+\mathcal{K})}\equiv-\frac{\mathcal{H}^2}{3{\phi^{\prime}}^2}.
\end{equation}
This equivalence can be shown easily by simply substituting the barotropic index for the scalar field and using the background equations (\ref{F1}) and (\ref{F2}) to simplify it further. The corresponding relation between $\zeta$ and the Mukhanov--Sasaki variable for the flat case is given and discussed in \cite{paper:GM}. 

\section{\label{sec:4}Quantization}
We intend to calculate the power spectrum for the slow-roll approximation for which the following conditions are satisfied:
\begin{equation}
    \dot{\phi}\ll\mathcal{V}(\phi), \quad  \ddot{\phi}\ll3 H\phi.
    \label{cond}
\end{equation}
The slow-roll parameters $\epsilon$ and $\delta$ are defined by
\begin{equation}
    \varepsilon=1-\frac{\mathcal{H}^{\prime}}{\mathcal{H}^{2}},\label{srp1}\quad
    \delta=\varepsilon-\frac{\varepsilon^{\prime}}{2\mathcal{H}\varepsilon}. \label{srp2}
\end{equation}
The background equations (\ref{F1}) and (\ref{F2}) when written in terms of slow-roll parameters take the form
\begin{eqnarray}
 &\frac{a^{\prime\prime}}{a}&={\mathcal{H}}^2(2-\epsilon) \label{F11},\\
 &{\mathcal{H}}^2&\Big(1-\frac{\varepsilon}{3}\Big)=\frac{2a^2}{m_{\rm P}^2}\mathcal{V}(\phi)-\frac{2\mathcal{K}}{3}.
 \label{F21}
\end{eqnarray}
The conformal time can be expressed in terms of slow-roll parameters by 
\begin{equation}
    \eta=-\frac{1}{\mathcal{H}}(1+\varepsilon)+\mathcal{O}(2),
    \label{eta}
\end{equation}
where the notation $ \mathcal{O}(2)$ is used to represent quadratic terms in $\varepsilon$ and $\delta$.

Applying the canonical quantization scheme, that is, taking the Hamiltonian constraint (\ref{H}) as an operator acting on the wave functional, leads to the Wheeler--DeWitt equation \cite{book:Kiefer}.\footnote{The formalism of canonical quantization is equivalent to the formalism of path integration; see, for example, Sec.~5.3.1 in \cite{book:Kiefer}.} Making a product ansatz for the full wavefunctional based on the assumption that perturbation modes do not interact with each other, we obtain for each mode the equation
\begin{equation}
\frac12\Bigg\{ e^{-2\alpha}\Bigg[ \frac{1}{m_{\rm P}^2}\frac{\partial^2}{\partial \alpha^2}+m_{\rm P}^2\Bigg( e^{6\alpha}{H}^2\Big(1-\frac{\varepsilon}{3}\Big)-\frac{\mathcal{K}}{3}e^{4\alpha}\Bigg)\Bigg]-\frac{\partial^2}{\partial {\tilde{v}_n}^2 }+\omega_n^2(\eta){\tilde{v}_n}^2 \Bigg\}\Psi_n\big(\alpha, \tilde{\phi},{\tilde{v}_n}\big) =0,
\label{WD}
\end{equation}
where we have rescaled $\phi$ to a dimensionless variable,
\begin{equation}
   \tilde{\phi}=m_{\rm P}^{-1}\phi.
\end{equation}
In the Wheeler--DeWitt equation we have ignored the kinetic term $\partial^2/\partial \phi^2$, which is the quantum analogue of neglecting the classical kinetic term as expressed by (\ref{cond}).

Note that a consistent quantization can be performed by using real variables which, however, is not the case for the variable $\tilde{v}_n$. Following, for example, \cite{paper:Martin}, we could construct a set of real variables for $\tilde{v}_n$ and $\tilde{p}_n$. But since such a redefinition would not affect our calculations, we will not introduce them explicitly and treat instead our variables as real variables.  

In terms of the slow-roll parameters, the frequency term $\omega_n^2(\eta)$ assumes the form
\begin{eqnarray}
  \omega_n^2(\eta)=&(n^2&-\mathcal{K})\Bigg(1+\frac{\mathcal{K\varepsilon}}{n^2-4\mathcal{K}}\Bigg)+\frac{3(\varepsilon+\delta)}{\eta^2}
  \Bigg[1+\frac{6\big(3-\varepsilon\big)}{n^2-4\mathcal{K}}\Bigg]-\frac{2+3\varepsilon}{\eta^2}\nonumber\\&+&\frac{1}{n^2-4}\Bigg\{-\frac{6\varepsilon}{\eta^2}(n^2-1) \Bigg(1-\frac32\frac{1}{n^2-4}\Bigg) -\mathcal{K}(1+6\varepsilon)\nonumber
\\
  &+&\mathcal{K}(n^2-1)\Bigg[-2(1+2\varepsilon)+\frac{3(3-2\varepsilon)}{n^2-4}\Bigg]
      \Bigg\},
      \label{omegansr}
\end{eqnarray}
where terms containing $\mathcal{K}/\mathcal{H}^2$ have been ignored. This approximation is valid as the comoving Hubble horizon $\mathcal{H}^{-1}$ decreases strongly during inflation. 

In the large-wavelength limit, we recover the result for the flat case, see e.g. \cite{book:Peter},
\begin{equation}
    \omega_n^2(\eta)=n^2-\frac{2+6\varepsilon-3\gamma}{\eta^2},
    \label{omegansrf}
\end{equation}
where we have introduced
\begin{equation}
    \gamma:=2\varepsilon-\delta.
\end{equation}
In the following, we shall apply a Born--Oppenheimer type of approximation scheme to the Wheeler--DeWitt equation. This allows us to recover first the dynamics of the classical background, then the Schr\"{o}dinger equation for the perturbations propagating on the classical background and, finally, quantum-gravitational corrections to the Schr\"odinger equation. The discussion of the corrections terms will be relegated to a future paper. Details for the general framework can be found, for example, in \cite{paper:SP1989,paper:KPS1991,book:Kiefer}. The formalism is flexible enough to be applicable to more general theories such as conformal gravity \cite{paper:KN2017}.  

We start by making a WKB-like ansatz, 
\begin{equation}
   \Psi_n\big(\alpha,\phi,{\tilde{v}_n}\big)=e^{i S(\alpha, \phi,\tilde{v}_n)},
   \label{wkba}
\end{equation}
and implement the Born--Oppenheimer approximation scheme by expanding\\ $S(\alpha, \phi,\tilde{v}_n)$ in terms of powers of the Planck mass,
\begin{equation}
 S(\alpha,\phi,\tilde{v}_n)=m_{\rm P}^2S_0+m_{\rm P}^0S_1+m_{\rm P}^{-2}S_2+ \ldots .  
\end{equation}
We then insert (\ref{wkba}) into the Wheeler--DeWitt equation (\ref{WD}) and derive equations at consecutive orders of $m_{\rm P}$.

At order $m_{\rm P}^2$, the Hamilton--Jacobi equation for the classical background is obtained, which takes the form
\begin{equation}
    \Bigg(\frac{\partial S_0}{\partial \alpha}\Bigg)^2-e^{6\alpha}H^2\Big(1-\frac{\varepsilon}{3}\Big)+\frac{\mathcal{K}}{3}e^{4\alpha}=0;
    \label{HJ}
\end{equation}
its solution reads
\begin{equation}
S_0(\alpha)=\pm \frac{1}{H^2(3-\varepsilon)}{\Bigg( H^2\Big(1-\frac{\varepsilon}{3}\Big)e^{2\alpha}-\frac{\mathcal{K}}{3}\Bigg)}^{3/2}.
\label{S0}
\end{equation}

At the next order, $m_{\rm P}^0$, we get an equation for $S_1$, which can be rewritten as a Schr\"{o}dinger equation for wave functions $\psi_n^{(0)}$,
\begin{equation}
     i\frac{\partial}{\partial \eta}\psi_n^{(0)}={^s{H_{|2}^n}}\psi_n^{(0)},
     \label{Se}
\end{equation}
where the perturbation Hamiltonian operator is 
\begin{equation}
{^s{H_{|2}^n}}=-\frac12\frac{\partial^2}{\partial {\tilde{v}_n}^2 }+\frac12\omega_n^2(\eta){\tilde{v}_n}^2;
\label{Hop}
\end{equation}
we have also introduced the conformal time parameter $\eta$ according to
\begin{equation}
    \frac{\partial}{\partial \eta}:=-e^{-2\alpha}\frac{{\partial}S_0}{\partial \alpha}\frac{{\partial}}{\partial{\alpha}}= e^{\alpha}\sqrt{ H^2\Big(1-\frac{\varepsilon}{3}\Big)-\frac{\mathcal{K}}{3}e^{-2\alpha}}\frac{{\partial}}{\partial{\alpha}},
\end{equation}
where we have fixed the sign of $S_0$ such that the time direction coincides with the expansion of the universe.

For the wave functions $\psi_n^{(0)}$ we make the following Gaussian ansatz:
\begin{equation}
 \psi_n^{(0)}(\eta,\tilde{v}_n)=A_n^{(0)}(\eta)e^{-\frac{1}{2}\Omega_n^{(0)}(\eta)\tilde{v}_n^2},
 \label{GA1}
\end{equation}
where $A_n^{(0)}(\eta)$ is fixed by the normalization of the wave function. 
Substituting this ansatz into the Schr\"{o}dinger equation (\ref{Se}), we obtain 
\begin{equation}
i \Omega_n^{\prime (0)}(\eta)= {\big(\Omega_n^{(0)}(\eta)\big)}^2-\omega_n^2.
\label{eqomega}
\end{equation}
To solve this equation, we use the following substitution:
\begin{equation}
 \Omega_n^{(0)}(\eta)=-i\frac{y_n^{\prime}(\eta)}{y_n(\eta)},  
 \label{sub}
\end{equation}
which leads to a second-order differential equation, the solution of which is given by
\begin{equation}
    y_n(\eta)=(-n \eta)^{1/2}\Big[c_{n,1}J_{-\nu_{\varepsilon,\delta}(n)}\big(-n\eta f_\varepsilon(n)\big)+c_{n,2}J_{\nu_{\varepsilon,\delta}(n)}\big(-n\eta f_\varepsilon(n)\big)\Big].
    \label{yn}
\end{equation}
Here,
\begin{equation}
f_\varepsilon(n):=\sqrt{1-\frac{1}{n^2}-\frac{1}{n^2-4}\Bigg({2+3\varepsilon}+\frac{3\varepsilon-1}{n^2}-\frac{3(3-2\varepsilon)}{n^2-4}\Big(1-\frac{1}{n^2}\Big)\Bigg)},
\label{f}
\end{equation}
and
\begin{equation}
    \nu_{\varepsilon,\delta}(n):=\big(\gamma+3/2\big)(1-\lambda_{\varepsilon,\delta}(n))
\end{equation}
\textnormal{with}
\begin{equation}
  \lambda_{\varepsilon,\delta}(n):=1-\sqrt{1-\frac{8}{n^2-4}\frac{1}{4\gamma+3}\Bigg(3\gamma+12\delta+\frac{3\varepsilon}{2}\frac{n^2-1}{n^2-4}\Bigg)}; \label{lambdan}
\end{equation}

The functions $y_n(\eta)$ are normalized according to 
\begin{equation}
    y_n^\prime y_n^*-y_n^{\prime *}y_n=i.
\end{equation}
We now choose for sub-Hubble modes the adiabatic vacuum state. In this state, the modes behave locally as in the Minkowski vacuum. It is defined by the asymptotic condition
\begin{equation}
        y_n(\eta) \rightarrow \frac{1}{\sqrt{2n}}e^{-in\eta}, \quad n \eta\rightarrow -\infty.
        \label{BD}
    \end{equation}
We thus have to set in (\ref{yn})
\begin{equation}
 c_{n,1}=-c_{n,2}e^{-i \pi\nu_{\varepsilon,\delta}(n)}, \quad  c_{n,2}=-\frac{i}{2}\sqrt{\frac{\pi f_\varepsilon(n)}{n}}\frac{e^{i n\eta(1-f_\varepsilon(n))} e^{-i\pi/4+i\pi\nu_{\varepsilon,\delta}(n)/2}}{\sin{(\pi\nu_{\varepsilon,\delta}(n))}}.   
\end{equation}  
At the next order, $m_{\rm P}^2$, we obtain correction terms for the Schr\"odinger equations (\ref{Se}). They will give tiny corrections to the power spectrum, similarly to the corrections for the flat case calculated in \cite{paper:David,paper:Manuel,paper:CK2021}. The discussion of these correction terms is beyond the scope of this paper and will be presented elsewhere.

\section{\label{sec:5}Power spectrum}

\subsection{General expression}

To derive the power spectrum, we have to calculate the two-point correlation function of $\tilde{v}_n$. 
It can be shown that (see e.g. \cite{paper:Martin})
\begin{equation}
     \big \langle \psi|\hat{\tilde{v}}_n\hat{\tilde{v}}_{n^{\prime}}^* |\psi\big \rangle=\frac{1}{2\mathcal{R}e \Omega_n^{(0)}}\delta_{n n^{\prime}}\delta_{l l^{\prime}}\delta_{m m^{\prime}},
    \label{corrf}
    \end{equation}
where the wave function $\psi$ is defined by $\psi=\prod_n\psi_n^{(0)}$.
We also have the following relation \cite{paper:Uzan},
\begin{equation}
    \big \langle \psi|\hat{\tilde{v}}_n\hat{\tilde{v}}_{n^{\prime}}^* |\psi\big \rangle= \frac{2 \pi^2}{n(n^2-\mathcal{K})}\mathcal{P}_{\tilde{v}}\delta_{n n^{\prime}}\delta_{l l^{\prime}}\delta_{m m^{\prime}}
    \label{corrP}
    \end{equation}
leading to the following expression for the power spectrum:
\begin{equation}
 \mathcal{P}_{\tilde{v}}= \frac{n(n^2-\mathcal{K})}{2 \pi^2}\frac{1}{2\mathcal{R}e \Omega_n^{(0)}}.
 \label{PowerS}
\end{equation}  
Using (\ref{sub}) and (\ref{yn}), we obtain in the super-Hubble limit $(-n\eta\rightarrow0)$ for the real part of $\Omega_n^{(0)}(\eta)$ the expression
\begin{equation}
  \mathcal{R}e\Omega_n^{(0)}(\eta)=\frac{\pi n f_\varepsilon(n)2^{-2\nu_{\varepsilon,\delta}(n)+1}}{\Gamma^2(\nu)}(-n\eta f_\varepsilon(n))^{2\nu_{\varepsilon,\delta}(n)-1}.
  \label{Reomega}
\end{equation}
Substituting this into (\ref{PowerS}) and applying further approximations using the smallness of the slow-roll parameters, we arrive at
\begin{equation}
 \mathcal{P}_{\tilde{v}}
  =\frac{n^2-{\mathcal{K}}}{4 \pi^2 f_\varepsilon^3(n)\xi^2}\Big[\tilde{C}_{\varepsilon,\delta}-2(\gamma-3\lambda_{\varepsilon,\delta}(n)/2)\ln\big(\xi f_\varepsilon(n)\big)\Big],
  \label{PS}
\end{equation}
where
\begin{equation}
  \tilde{C}_{\varepsilon,\delta}=1-2\varepsilon+(\gamma-3\lambda_{\varepsilon,\delta}(n)/2)(4-2\gamma_E-2\ln(2)) \ \ \textnormal{and} \ \ \xi=\frac{n}{a H},
  \label{Ctilde}
\end{equation}
and $\gamma_E\simeq0.5772$ is the Euler--Mascheroni constant.

In the large-wavelength limit, we recover the result for the flat case \cite{paper:Manuel},
\begin{equation}
 \mathcal{P}_{{v}}= \frac{k^2}{4 \pi^2\xi^2}\Big[ C_{\varepsilon,\delta}-2\gamma\ln\big(\xi\big)\Big],
\end{equation}
where
\begin{equation}
    C_{\varepsilon,\delta}=1-2\varepsilon+\gamma(4-2\gamma_E-2\ln(2));
\end{equation}
see also \cite{book:Peter}, p.~498.
Furthermore, we need to calculate the power spectrum for the parameter $\zeta_{\rm BST}$. In the super-Hubble limit, the second component in (\ref{zetaBSTphi}) can be ignored. Hence, for the slow-roll regime, the relation (\ref{zetaBSTphi}) reads as follows:
\begin{equation}
 \zeta_{{\rm BST},n}=-\frac{1}{a}\frac{1}{\sqrt{2\left(\varepsilon+\frac{\mathcal{K}}{\mathcal{H}^2}\right)}M_{\rm P}}\Bigg[1+\frac{\mathcal{K}}{n^2-4\mathcal{K}}\left(\varepsilon+\frac{\mathcal{K}}{\mathcal{H}^2}\right)\Bigg]^{1/2}\tilde{v}_n,
 \end{equation}
where we have introduced a redefined Planck mass $M_{\rm P}:=1/\sqrt{8\pi G}$.
With this, the power spectrum of $\zeta_{\rm BST}$ takes the form
\begin{equation}
    \mathcal{P}_{\zeta_{\rm BST}}= \frac{1}{2a^2\left(\varepsilon+\frac{\mathcal{K}}{\mathcal{H}^2}\right)M_{\rm P}^2}\Bigg[1+\frac{\mathcal{K}}{n^2-4\mathcal{K}}\left(\varepsilon+\frac{\mathcal{K}}{\mathcal{H}^2}\right)\Bigg] \mathcal{P}_{\tilde{v}_n}.
    \label{PzBST}
\end{equation}
Let us emphasize here that the power spectrum has no trivial scale (or wavelength) dependence. As we show in the next section, particularly for the large scales where we see the suppression of the spectrum, it does not behave as a power-law.

Here again, considering the large-wavelength limit, we obtain 
\begin{equation}
    \mathcal{P}_{\zeta}= \frac{1}{2a^2\varepsilon M_{\rm P}^2}\mathcal{P}_{v_n},
    \label{Pz}
\end{equation}
which is the result for the flat case \cite{paper:Manuel}. (The tensor modes, which are not discussed here, can be obtained by setting $\gamma=\epsilon$.)
\begin{figure}
\centering
\includegraphics[width=0.5\textwidth]{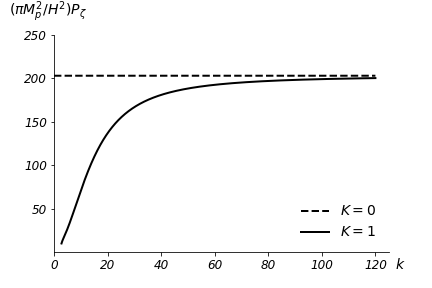}
\caption{Plot of the power spectra of curvature perturbations (multiplied by $\pi M_{\rm P}^2/H^2$) for the flat (dashed line) and closed (thick line) models of a universe at the point of horizon reentry, i.e. $\xi=1$. The slow-roll parameters are set to $\varepsilon=0.005$ and $\delta=-0.006$.}\label{PlotPz}
\end{figure}

In Fig.~\ref{PlotPz}, the power spectra of curvature perturbations are plotted as a function of the comoving wavelength for the flat and the closed model. One can see that, compared to the flat case, there is a significant suppression of power at large scales for the closed case. One can also recognize from the plot the natural cutoff at small $k$ due to the finite radius of this universe. 

Let us recall that the curvature radius of a closed model of the universe is given by $R_{\rm c}=\frac{c}{H_0}\frac{1}{\sqrt{|\Omega_k|}}$ and that the following relation holds: $n^2=\mathcal{K}(1+R_{\rm c}^2 k_{{\rm phys}}^2)$, where $k_{{\rm phys}}$ is the physical wavelength \cite{Efst}. We can make the following rough estimate. Taking the Hubble parameter to be $67.4$ km/(s Mpc) and $\Omega_k\approx -0.037$ \cite{paper:Planck2018}, one can estimate to have suppression at a scale of approximately $3.6$ Gpc and a cut at around $52$ Gpc. An exact calculation can be made only by numerical means.

\subsection{\label{sec:5.2}Power-law approximation}
The power spectrum for a single-field slow-roll model of inflation can be approximated by a power-law spectrum \cite{book:Peter}, which is in agreement with observations \cite{paper:Planck2018},
\begin{equation}
\ln{\mathcal{P}_\zeta}=\ln{\mathcal{P}_\zeta(k_*)}+[n_s(k_*)-1]\ln{\frac{k}{k_*}}+\frac12\alpha_s \ln^2{\frac{k}{k_*}}+ \ldots ,   
\end{equation}
where $k_*$ is the pivot scale, $A_s\equiv\mathcal{P}_{\zeta}(k_*)$ is called a scalar power spectrum amplitude, $n_s$ is the spectral index, and $\alpha_s$ is the  running of the scalar index. This approximation is based on the assumption that there is only a weak scale dependence which can be quantified by the spectral index.

From (\ref{PzBST}), it is obvious that there is an explicit scale dependence that enters due to the curvature of the universe. We will thus not be able to approximate this power spectrum by a power-law at large scales where the scale dependence is prominent. Indeed, calculating the spectral index for $\mathcal{K}=1$ case, we arrive at
\begin{eqnarray}
     n_s-1\approx&-&2\varepsilon-\frac{2\varepsilon(\varepsilon-\delta)}{\varepsilon+1/n^2}+\frac{2}{n^2}\Bigg(\frac{1}{\varepsilon+1/n^2}+\frac{1}{1-1/n^2}\Bigg)\nonumber
     \\&
     -&\frac{n}{n^2-4+\varepsilon+1/n^2}\Bigg[\frac{2 n }{n^2-4}\left(\varepsilon+\frac{1}{n^2}\right)+\frac{2}{n^3}\Bigg]\nonumber
     \\&-&\frac{\ln{[f_\varepsilon(n)]}}{1-\lambda_{\varepsilon,\delta}(n)}\frac{12 n}{n^2-4}\frac{1}{4\gamma+3}\Bigg\{(3\gamma+12\delta)\left[\frac{2n}{n^2-4}-1\right]+3\varepsilon \frac{n(n^2-2)}{(n^2-4)^2}\Bigg\}\nonumber
     \\&
     -&\frac{{2\gamma-3\lambda_{\varepsilon,\delta}(n)+3}}{f_{\varepsilon}^2(n)}\Bigg\{ \frac{1 }{n^2}+\frac{ n^2}{(n^2-4)^2}\Bigg[2+3\varepsilon
     \nonumber
     \\&
     +&\frac{3\varepsilon-1}{n^2}\Bigg(2-\frac{4}{n^2}\Bigg)-\frac{6(3-2\varepsilon)}{n^2-4}\left(1-\frac{1}{n^2}-\frac{n^2-4}{2n^4}\right)\Bigg]\Bigg\}.  
\end{eqnarray}
We note that in the large-wavelength limit the result for the flat model is obtained \cite{book:Peter},
\begin{equation}
  n_s-1=2\delta-4\varepsilon.
\end{equation}
We thus explicitly recognize that for large scales, where the main deviation from the flat model occurs, the power-law approximation is not applicable. 
The next step in comparing our results with observations would consist in  determining the angular power spectrum using numerical tools; see, for example, \cite{paper:LT} or the Numerical Cosmology library – NumCosmo \cite{code:num}, in order to fit the theoretical result to the Planck data. This is relegated to future work. 

\section{\label{sec:6}Conclusion}

The question whether our Universe is open or closed is among the most important open questions in cosmology. Observations indicate that it is spatially flat, but this is not without controversy \cite{paper:Valentino,paper:VLM}. There are also convincing conceptual and mathematical arguments that suggest that space cannot be infinitely large \cite{paper:EMN}. These points have motivated us to have a fresh look at a closed universe and to calculate the power spectrum for scalar modes in a slow-roll model of inflation from a fundamental point of view. This means that we started from a theory of quantum gravity (quantum geometrodynamics) and derived the power spectrum using a Born--Oppenheimer type of scheme for the wave function. We made use of gauge-invariant variables throughout. Our main result is the analytical expression (\ref{PzBST}) for the power spectrum of the curvature perturbations. It describes a suppression of power at large scales. The angular power spectrum can only be calculated by sophisticated numerical methods, which is beyond the scope of this paper.
By employing a closed model, one might be able to explain the observed lack of power for large scales \cite{paper:Planck2018}.

It must be emphasized that other, unrelated, models can also lead to a suppression of power at large scales. One example is fast-roll inflation for an open (${\mathcal K}=-1$) model \cite{WZS14}. There, the suppression can be traced directly to a period of fast roll during inflation. The treatment in our paper relies on slow-roll inflationary models, which are the most common ones. A generalization to fast-roll models is beyond the scope of our paper.

In our paper, we have not considered the power spectrum for the tensorial modes. In order to achieve this, one should start with the tensorial part of the perturbation Hamiltonian, bring it into a gauge-invariant form and follow the steps performed here for the scalar perturbations. For the flat case, there is a simple relation between scalar and tensor perturbations, but obtaining an analogous relation for the closed case (which is anticipated to be more complicated compared to the flat case), can be realized only after a significant amount of calculations. For this reason, the calculation of the tensorial power spectrum is beyond the scope of this paper and relegated to future work. 

Our formalism is suitable to calculating genuine quantum-gravitational effects. This can be achieved by proceeding with the Born--Oppenheimer approximation to higher orders in the inverse Planck-mass squared. In the flat case, the power spectrum was calculated along these lines in \cite{paper:David,paper:Manuel,paper:excited,paper:CK2021}. 
So far, these terms are too tiny to be observable, but they constitute concrete predictions from a concrete approach to quantum gravity and may become relevant for future applications. The calculations of such terms for closed models is left for future investigation. 

\begin{acknowledgements}
 We thank Patrick Peter for helpful discussions and comments.
\end{acknowledgements}

\vskip 5mm

\noindent {\bf Data Availability Statement}: Data sharing not
applicable--no new data generated.

\vskip 5mm

\noindent This paper is published open access in {\em General Relativity
  and Gravitation} {\bf 54}, Article number: 30 (2022),
https://doi.org/10.1007/s10714-022-02918-3. 

\newpage

\appendix

\section{\label{app:sec1}Harmonics on the three-sphere}
We give here a brief introduction to scalar, vector and tensor
harmonics on the three-sphere following \cite{paper:Abbott}, \cite{paper:Riaz}, \cite{paper:Hawking}, and \cite{paper:Gerlach}.\\
We write the FLRW metric as
\begin{equation}
     d s^2=a^2(\eta)\left[-d\eta^2+d \chi^2+ s^2_{\mathcal{K}}(\chi) d \Omega^2 \right],
 \end{equation}
where $\chi$ is the comoving radial distance, the infinitesimal solid angle $d \Omega^2=d \theta^2+ \sin^2{\theta} d \phi^2 $ and
$$
s_{\mathcal{K}}(\chi) = \left\{ \begin{array}{ll}
\sinh{\left(\sqrt{|\mathcal{K}|}\chi\right)}/\sqrt{|\mathcal{K}|},  &\mbox{ $\mathcal{K}=-1,$} \\
 \chi, &\mbox{ $\mathcal{K}=0,$}\\
\sin{\left(\sqrt{|\mathcal{K}|}\chi\right)}/\sqrt{|\mathcal{K}|},   &\mbox{ $\mathcal{K}=+1,$}
       \end{array} \right.
$$

\subsection{\label{app:subsec1}Spherical harmonics}
Scalar harmonics are solutions of a generalized Helmholtz equation. 
In a $D$-dimensional maximally symmetric space the Helmholtz equation is written as
\begin{equation}
    D^2 Q^{(n)}(\chi, \theta,\phi)=-k^2\ Q^{(n)}(\chi, \theta,\phi),
\end{equation}
where $D^2=D^i D_i$, and $D_i$ is a spatial covariant derivative. The comoving wavenumber $k$ and the eigenmode $n$ are related by $k^2=n^2-\mathcal{K}$, with 
$$
k^2 = \left\{ \begin{array}{lll}
 n^2+1, &\mbox{ $n\geq0,$}  &\mbox{ $\mathcal{K}=-1,$} \\
  n^2, &\mbox{ $n\geq0,$}  &\mbox{ $\mathcal{K}=0,$}\\
 n^2-1, &\mbox{ $n=1,2,3,...,$}  &\mbox{ $\mathcal{K}=+1.$}
       \end{array} \right.
$$
The general solution $Q^{(n)}(\chi, \theta,\phi)$ can be written as a sum of scalar spherical harmonics $Q^{n}_{lm}(\chi, \theta,\phi)$, which can be decomposed into radial and angular parts, 
\begin{equation}
    Q^{n}_{lm}(\chi, \theta,\phi)=\Pi^{n}_{l}(\chi)Y_{l m}(\theta, \phi),
\end{equation}
where $Y_{l m} (\theta, \phi)$ are the spherical harmonics.
The radial eigenfunctions $ \Pi^{n}_{ l}(\chi)$ are solutions of the radial harmonic equation and are given by
$$
 \Pi^{n}_{ l}(\chi) = \left\{ \begin{array}{ll}
\sqrt{\frac{N_{n l}}{s_{\mathcal{K}}(\chi)}}P_{-1/2+i n}^{-1/2-l}(\cosh{(\sqrt{-\mathcal{K}}\chi)}),  &\mbox{ $\mathcal{K}=-1,$} \\
 \sqrt{\frac{2 k^2}{\pi}}j_l(k r), &\mbox{ $\mathcal{K}=0,$}\\
\sqrt{\frac{M_{n l}}{s_{\mathcal{K}}(\chi)}}P_{-1/2+n}^{-1/2-l}(\cos{(\sqrt{\mathcal{K}}\chi)}),   &\mbox{ $\mathcal{K}=+1,$}
       \end{array} \right.
$$
with the coefficients 
\begin{equation}
    N^{n}_{ l} = \prod _{p=0}^{l} (n^2+p^2)\ \ \ \textnormal{and}\  \ \
    M^{n}_{ l} =  \prod _{p=0}^{l} (n^2-p^2).
\end{equation}
We use the normalization condition 
\begin{equation}
    \int   Q^{n *}_{l m}(\chi, \theta, \phi) Q^{n^{\prime}}_{ l^{\prime} m^{\prime}}(\chi, \theta, \phi)s^2_{\mathcal{K}}(\chi) d \chi d\Omega  =\delta(n,n^{\prime})\delta_{l l^{\prime}}\delta_{m m^{\prime}},
\end{equation}
where
$$
\delta(n,n^{\prime})= \left\{ \begin{array}{ll}
\delta(n-n^{\prime})=\delta(k-k^{\prime}),  &\mbox{ $\mathcal{K}=0,-1,$} \\
\delta_{n n^{\prime}},   &\mbox{ $\mathcal{K}=+1.$}
       \end{array} \right.
$$
\subsection{\label{app:subsec2}Vector harmonics}
The vector harmonics $(S_i)^n_{l m}(\chi,\theta, \phi)$ are vector solutions of the generalized Helmholtz equation 
\begin{equation}
 D^2 S_i^{(n)}=-k^2 S_i^{(n)}, \ \ \ \ \ n=2,3,4, \ldots ,
\end{equation}
where $n^2=k^2+2\mathcal{K}$ and $S_i^{(n)}$ are transverse, i.e. the following condition is satisfied:
\begin{equation}
    D^i S_i^{(n)}=0.
\end{equation}
Here again, the general solution can be written as a sum of  $(S_i)^n_{l m}(\chi,\theta,\phi)$ vector harmonics.

We use the following normalization condition
\begin{equation}
    \int d\mu (S_i)_{l m}^n(S^i)_{l^{\prime}m^{\prime}}^{n^{\prime}}=\delta^{n n^{\prime}}\delta_{l l^{\prime}}\delta_{m m^{\prime}};
\end{equation}
using a parity transformation,
the $(S_i)^n_{l m}$-harmonics can be decomposed into linearly independent even, $(S_i^e)^n_{l m}$, and odd, $(S_i^o)^n_{l m}$, components (see the detailed explanation in \cite{paper:Gerlach}). 

The complete orthogonal set is composed of three vector harmonics: $(S_i^e)^n_{l m}$, $(S_i^o)^n_{l m}$, and $(P_i)^n_{l m}$. The last one can be constructed via the scalar harmonics $Q^n_{l m}$, 
\begin{equation}
  (P_i)^n_{l m}=\frac{1}{n^2-\mathcal{K}}D_i Q^n_{l m}, \ \ \ \ n=2,3,4, \ldots ,
\end{equation}
which satisfies
\begin{equation}
    D^2 (P_i)^n_{l m}=-(n^2-3\mathcal{K}) (P_i)^n_{l m} \ \ \ \textnormal{and} \ \ \ D^i (P_i)^n_{l m}=-Q^n_{l m}.
    \end{equation}
The normalization condition is
\begin{equation}
    \int d\mu (P_i)_{l m}^n(P^i)_{l^{\prime}m^{\prime}}^{n^{\prime}}=\frac{1}{n^2-\mathcal{K}}\delta^{n n^{\prime}}\delta_{l l^{\prime}}\delta_{m m^{\prime}}.
\end{equation}

\subsection{\label{app:subsec3}Tensor harmonics}
The tensor harmonics $(G_{i j})_{l m}^n(\chi, \theta,\phi)$ are tensor solutions of the generalized Helmholtz equation 
\begin{equation}
    D^2 G_{i j}^{(n)}=-k^2 G_{i j}^{(n)}, \ \ \ \ n=3,4,5, \ldots ,
\end{equation}
where $n^2=k^2+3\mathcal{K}$. 

The tensor harmonics are transverse and traceless; therefore, 
\begin{equation}
    D^i G_{i j}^{(n)}=0, \ \ \ \ \ \ G_{i}^{(n)i}=0.
\end{equation}
They can again be classified as even $(G_{i j}^e)^{n}_{l m}$ and odd $(G^o_{i j})^{n}_{l m}$ linearly independent harmonics.

The normalization condition is
\begin{equation}
    \int d\mu (G_{i j})_{l m}^n(G^{i j})_{l^{\prime}m^{\prime}}^{n^{\prime}}=\delta^{n n^{\prime}}\delta_{l l^{\prime}}\delta_{m m^{\prime}}.
\end{equation}
The complete orthogonal set is composed of six tensor harmonics: $Q_{i j}$, $P_{i j}$, $S_{i j}^o$, $S_{i j}^e$, $G_{i j}^0$, and $G_{i j}^e$. The first two, i.e., $Q_{i j}$ and $P_{i j}$, are constructed using the scalar  harmonics and  $S_{i j}^o$, $S_{i j}^e$ are constructed via odd $(S_i^o)^n_{l m}$, and even $(S_i^e)^n_{l m}$ transverse vector harmonics (see \cite{paper:Hawking}). Note that for $\mathcal{K}=+1$, the mode values $n=1$ and $n=2$ should not be included because expanding functions in harmonics on the three-sphere as certain tensor harmonics is impossible for those values \cite{Lifshitz}.

\section{\label{app:sec2}Perturbations of the metric and the scalar field}
Following \cite{paper:Hawking}, the perturbations of the line element (\ref{ADM}) are considered. For convenience, we use the same notation as in \cite{paper:Hawking}.
The perturbation of the three-metric $ \epsilon_{ij}$ when expanded in harmonics reads
\begin{eqnarray}
    \epsilon_{ij}=\sum_{n,l,m}\Big[&6^{1/2}&a_{nlm}\frac13\gamma_{ij}Q_{lm}^n+6^{1/2}b_{nlm}(P_{ij})^n_{lm}+2^{1/2}c_{nlm}^o(S_{ij}^o)^n_{lm}\nonumber\\&+&2^{1/2}c_{nlm}^e(S_{ij}^e)^n_{lm}+2^{1/2}d_{nlm}^0(G_{ij}^0)^n_{lm}+2^{1/2}d_{nlm}^e(G_{ij}^e)^n_{lm}\Big].
    \label{epsilon}
\end{eqnarray}
The lapse function $N$, shift vector $N_i$, and scalar field $\Phi_{\rm f}$ are expanded in terms of harmonics as
\begin{eqnarray}
    &N&=N_0\Big[1+6^{-1/2}\sum_{n,l,m}g_{nlm}Q^n_{lm}\Big],\label{Nsh}\\
    &N_i&=a\sum_{n,l,m}\Big[6^{-1/2}k_{nlm}(P_i)^n_{lm}+2^{1/2}j_{nlm}(S_i)_{lm}^n\Big],\label{Nish}\\
    &\Phi_{\rm f}&=\frac{1}{2^{1/2}\pi}\phi(t)+\sum_{n,l,m}f_{nlm}Q^n_{lm}.\label{phish}
\end{eqnarray}
The coefficients $a_{nlm}$, $b_{nlm}$, $c^o_{nlm}$, $c^e_{nlm}$, $d^o_{nlm}$, $d^e_{nlm}$ are functions of time. For brevity, as a label we write $n$ instead of labels $\{n,l,m\}$.

\section{\label{app:sec3}The gauge-invariant variable in terms of the Bardeen potential}
The goal of this section is to express the functions $f_n$ in terms of the Bardeen potential $\Phi$. This is necessary to ultimately obtain an expression for the gauge-invariant variable $\tilde{v}_n$ in terms of the Bardeen potential.
To accomplish this task, we need to use the field equations obtained by the variation of the action with respect to the perturbations.
We write the total action of the inflationary FLRW universe as \cite{paper:Hawking}
\begin{equation}
    S=S_0(\alpha,\phi, N_0)+\sum_n S_n,
    \label{Saction}
\end{equation}
where $S_0$ and $S_n$ are, respectively, the unperturbed (or background) and perturbation actions. The perturbation action can be written as
\begin{equation}
    S_n=\int d t (L_g^n+L_m^n),
    \label{Sn}
\end{equation}
where $L_g^n$ is the Lagrangian for the gravitational part and $L_m^n$ respectively for the matter part; see the explicit expressions of these Lagrangians in Appendix B of \cite{paper:Hawking}.

Working in longitudinal gauge and using (\ref{r1}) and (\ref{r2}), the perturbation Lagrangians\footnote{We perform the calculations by setting the units equal to one, and recover them in the final results only.} are written in terms of the Bardeen  potential,
\begin{eqnarray*}
    L_g^n=\frac{1}{2}a^2\left\{ \left(-2n^2-7-33{\mathcal{H}}^2\right) \Phi_n^2-6{\Phi_n^{\prime}}^2-24\mathcal{H}\Phi_n \Phi_n^{\prime}\right\}
    \label{Lgn}
\end{eqnarray*}
and
\begin{eqnarray*}
    L_m^n=\frac12 a^4 \Bigg\{&\frac{1}{a^2}&\left({f_n^{\prime}}^2-6\Phi_n f_n^{\prime}\phi^{\prime}\right)-m^2\left(f_n^2-6\Phi_n f_n\phi\right)-\frac{1}{a^2}(n^2-1)f_n^2+\frac{6{\phi^{\prime}}^2}{a^2}\Phi_n^2 \nonumber\\&+&9\left(\frac{{\phi^{\prime}}^2}{a^2}-m^2\phi^2\right)\Phi_n^2-\Phi_n\left[2m^2f_n\phi-3\cdot m^2\phi^2\Phi_n+2\frac{f_n^{\prime}\phi^{\prime}}{a^2}\frac{3{\phi^{\prime}}^2}{a^2}\Phi_n\right]\Bigg\}.
    \label{Lmn}
\end{eqnarray*}
Variation of the action $S_n$ with respect to $a_n$, $f_n$, and $g_n$ leads to two field equations and a constraint, which when written in terms of Bardeen potential are
 \begin{eqnarray}
&\Phi_n^{\prime\prime}+3\mathcal{H}\Phi_n^{\prime}+\left[3a^2m^2\phi^2-\frac13(n^2+2)+\frac{1}{3}(n^2-4)\right]\Phi_n=3 ({\phi}^{\prime}f_n^{\prime}-a^2m^2\phi f_n), \label{phiB}\\
&  f_n^{\prime \prime}+ 2  \mathcal{H} f_n^{\prime}+[m^2a^2+(n^2-1)]f_n=\Big[-2a^2m^2\phi \Phi_n+4\phi^{\prime}\Phi_n^{\prime}\Big] \label{fne}
\end{eqnarray}
and the constraint
\begin{equation}
2\phi^{\prime}f_n^{\prime}+2m^2a^2\phi f_n=\left[5 (-\mathcal{H}^2+{\phi^{\prime}}^2)+3m^2a^2\phi^2-\frac23\Big(n^2+\frac12\Big)\right]\Phi_n-2\mathcal{H}\Phi_n^{\prime}.\label{fnc}
\end{equation}
Using the above set of equations, the function $f_n$ can be given in terms of Bardeen potential,
\begin{eqnarray}
f_n=-\frac{1}{2\phi^{\prime}(n^2-1)}\Bigg\{&\Big[&-\mathcal{H}^2-2\mathcal{H}^{\prime} -3{\phi^{\prime}}^2+3m^2a^2\phi^2-\frac23\Big(n^2+\frac12\Big)\Big]\Phi_n^{\prime}\nonumber\\&+&\mathcal{H}\left[5a^2m^2\phi^2+5 (-\mathcal{H}^2+{\phi^{\prime}{}^2})-\frac23n^2-4-\frac{1}{3}\right]\Phi_n\Bigg\}.
\label{fn}
\end{eqnarray}
Substituting this into (\ref{Qn}), and using the background equations (\ref{F1}), (\ref{F2}), (\ref{KG}) to simplify it further, the following expression is obtained for the gauge-invariant variable $Q_n$:
\begin{equation}
    Q_n=\frac{\phi^{\prime}}{\mathcal{H}}\Phi_n+f_n=\frac{\mathcal{H}}{3\phi^{\prime}}\left\{{\mathcal{H}^{-1}}\Phi_n^{\prime}+\left[1+\frac{3\phi^{{\prime}^2}}{\mathcal{H}^2}\right]\Phi_n \right\}.
    \label{QnPhi}
\end{equation}

\end{document}